\documentstyle[preprint,revtex]{aps}
\begin{document}
\draft
\begin{title}
\begin{center}
The Hubbard Model at Infinite Dimensions:\\
Thermodynamic and Transport Properties
\end{center}
\end{title}
\author{Th.\ Pruschke, D.L.\ Cox}
\begin{instit}
Department of Physics, The Ohio State University,\\
Columbus, Ohio 43210-1106
\end{instit}
\author{M.\ Jarrell}
\begin{instit}
Department of Physics, University of Cincinnati,\\
Cincinnati, Ohio 45221-0011
\end{instit}
\receipt{May, 1992}
\begin{abstract}
We present results on thermodynamic quantities, resistivity and optical
conductivity for the Hubbard model on a simple hypercubic lattice in infinite
dimensions. Our results for the paramagnetic phase display the features
expected from an intuitive analysis of the one-particle spectra and
substantiate the similarity of the physics of the Hubbard model to those
of heavy fermion systems.

The calculations were performed using an approximate solution
to the single-impurity Anderson model, which is the key quantity entering
the solution of the Hubbard model in this limit. To establish the quality
of this approximation we compare its results, together with those obtained
from two other widely used methods, to essentially exact quantum Monte
Carlo results.
\end{abstract}
\narrowtext
\section{Introduction}
The limit of infinite spatial
dimensions has turned out to be a natural starting point for obtaining sensible
approximate \cite{mv89,bm,muha89-1,janis}, and even essentially
exact \cite{ja92,I} solutions of models of highly correlated electronic
systems.  In this limit the dynamics
of the system become essentially local \cite{muha89-1} which considerably
simplifies the task of calculating quantities of interest \cite{bm,ja92,gk,I}.

In the present paper we want to extend our previous study of the Hubbard
Hamiltonian
\cite{ja92,I,hbd}
\begin{equation}
H =\begin{array}[t]{l}
\displaystyle\sum_{\langle ij\rangle,\sigma}t_{ij} \left( c^\dagger_{i,\sigma}
c_{j,\sigma} + {\rm h.c.}\right)
-\mu\sum_{i\sigma}n_{i\sigma}\\[5mm]
\displaystyle+U \sum_i n_{i,\uparrow}n_{i,\downarrow}
\end{array}
\end{equation}
in the limit of infinite spatial dimensions $d=\infty$.
The notation in (1) is the standard one and the limit $d\to\infty$ has to be
taken
such that ${t^*}^2\equiv d\langle t_{{\bf k}}^2\rangle_{{\bf k}}={\rm
constant}$.
In \cite{ja92} one of us demonstrated that the one-particle Green's function,
or equivalently the proper one-particle self energy of the model (1), in
this limit is obtained from the equation
\begin{equation}
G_{ii}(z)=\int d\omega A_0(\omega)\frac{1}{z-\omega-\epsilon-\Sigma(z)}
\stackrel{!}{=}{\cal G}(z)\,.
\end{equation}
Here, the Green's function ${\cal G}(z)$ is the solution of a single-impurity
Anderson model with an effective hybridization given by \cite{I}
\begin{equation}
\Delta(z)=\frac{1}{G_{ii}(z)}+\Sigma(z)-z-\mu
\end{equation}
and $A_0(\epsilon)$ denotes the free one-particle density of states (DOS). Note
that for a given site $i$ equation (3) defines an effective potential due to
the presence of the lattice. Equations (2) and (3) thus constitute the
``natural'' mean field theory for the Hubbard model (1) \cite{janis}.

This mean field theory is of course independent of the lattice structure.
For reasons of convenience, however,  we shall concentrate on a
simple hypercubic lattice with $N$ sites and transfer along the $d$-coordinate
axes only, i.e.\ $\displaystyle t_{{\bf k}}=-2\sum_{m=1}^\infty
t_m\sum_{n=1}^d\cos(mk_n)$.
The latter assumption obviously oversimplifies the situation when one wants to
consider transfer beyond nearest neighbours, but it has the advantage
that the free single-particle DOS
\begin{equation}
A_0(\epsilon)=\frac{1}{N}\sum_{{\bf k}}\delta(\epsilon-t_{{\bf k}})
\end{equation}
acquires the simple Gaussian form \cite{muha89-1}
\begin{equation}
A_0(\epsilon) = \exp(-\epsilon^2)/\sqrt{\pi}
\end{equation}
when ${t^*}^2=\sum {t^*_m}^2=1$. The latter convention will set the energy
scale used for the remainder of this paper. For nearest-neighbour transfer
and ${\bf q}= (\pi,\pi,\ldots)$ one has perfect nesting. However,
any $t_2\not=0$ destroys
this property and thus allows us to continuously bias quantities which
depend on the perfect nesting like magnetic instabilities.

The situation with nearest-neighbour transfer only was explored in refs.\ 5\&7
using a quantum Monte Carlo (QMC) method to solve the impurity Anderson model.
We could thus obtain essentially exact results for the model (1) and discuss
magnetic and single-particle properties for a variety of model parameters and
temperatures. The results at half filling $n_e=1$
can most conveniently be presented in the phase diagram in Fig.~1: For small
values of $U$ and high temperatures one finds a paramagnetic metal with
correlation enhanced Fermi-liquid parameters. By increasing $U$ for a fixed
temperature
a crossover through a semi-metallic like (shaded region) into a Mott-Hubbard
like phase with exponentially reduced DOS at $\mu$
takes place. Note, that one never finds a true gap in the DOS for this
``phase''. Nevertheless, transport and thermodynamic properties will
essentially behave like an insulator. By lowering the temperature for
fixed $U$, one encounters an antiferromagnetic transition which is connected
with a gap in the one-particle DOS due to the cell-doubling associated with
the antiferromagnetic state. As mentioned earlier, this phase
can be shifted to lower temperatures or even be completely suppressed
by magnetically frustrating the system by a finite $t_2$. In this case the
dotted line in the antiferromagnetic region in Fig.~1 becomes important. It
visualizes the behaviour of the ``MI''-crossover for a fixed $U$ when the
temperature is lowered and shows an interesting and unexpected reentrance
behaviour. As we will discuss later, this is connected to a competition between
the ``Mott-Hubbard'' phase and the Kondo effect also present in this model
\cite{I}.

This Kondo effect is apparent as the temperature approaches some small
energy scale.  There are two ways we may define such an energy
scale.  First, we can obtain an ``effective impurity'' Kondo-scale
$T_K$ from the self-consistently embedded impurity by calculating the screened
local moment on the site $i$, $T\chi(T)_{ii}$, and then extracting $T_K$ by
fitting this result to the universal numerical renormalization group results
of Krishnamurthy et al. \cite{krish}.  We find that $T_K$ defined in this way
is strongly temperature dependent, increasing with decreasing $T$ \cite{I}.
This
temperature dependence  of $T_K$ can easily be understood --
and is in fact expected -- from the obvious temperature dependence of the
effective
medium defined by (3). It also explains the observed rather fast disappearance
of the Abrikosov-Suhl resonance (ASR) in the one-particle DOS, associated with
this Kondo screening,
with increasing temperature and its total absence in the semi-metallic
region of the phase diagram. Second, we may also identify
a scale $T_0$ as where the ASR in the density
of states reaches half its maximum value.  This latter energy scale
appears to be more physically meaningful since it shows up in physical
quantities
like specific heat and resistivity.

Away from half filling the Mott-Hubbard phase is immediately replaced by a
Fermi
liquid with strongly enhanced Fermi liquid parameters. More precisely, we found
a narrow resonance at $\mu$ for low temperatures which leads to the observed
enhanced quasiparticle
mass. This resonance could again be traced to a Kondo-screening of the local
moments with a dynamically generated low temperature energy scale -- $T_0$ --
connected to it. The
magnetic transition is also found to be suppressed  upon doping.
Finally, for greater than $20$\% doping, correlation effects become
less important and the system basically behaves like one would expect from
standard
perturbation theory.

The remainder of this paper is split into three parts.
First, we will compare different
approximation schemes to the QMC results. The main reason is that QMC is
rather time intensive and becomes problematic for large values of $U$ and
inverse temperature $\beta$. Also, by virtue of the
method, the QMC  process gives all dynamical quantities as function of
Matsubara rather than real frequency and one has to use e.g.\
maximum entropy methods to analytically continue these results to real
frequencies. Although
this is straightforward for densities of states, it proves
problematic for quantities like the one-particle self energy.
On the other hand, several physical quantities need this real frequency
dependence as input. As we will show, a good approximation scheme for this
purpose is given by a self-consistent perturbation theory developed for the
single-impurity Anderson model (NCA) \cite{cbw,pg89}. In the second part of the
paper we use this approximation to  calculate free energy, specific heat,
resistivity and optical conductivity for the model (1) for the
paramagnetic phase. Finally, a discussion
will close the paper.

\section{Comparison of different methods}
One major problem in using the QMC approach to calculate physical quantities
is the rather large amount of computer time one has to invest to obtain
results for one particular set of parameters. Especially for thermodynamic
properties,
where one has to adjust the chemical potential to maintain a fixed filling, it
is difficult to calculate a temperature series. It is thus
clearly desirable to have some different methods to solve the Hamiltonian
(1) or, equivalently, the single-impurity Anderson model.

The most straightforward idea is to use standard perturbation theory in $U$.
This is known to work rather well for the symmetric single-impurity Anderson
model \cite{zh,yy},
and one thus may expect it to be a reasonable approximation at half filling and
for
small values of $U$. Away from the symmetric point it is known that at least
the lowest order does not reproduce the correct occupation number
\cite{gk}. Nevertheless it is a simple method and it
is surely worthwhile to outline its region of applicability. It also has the
advantage that it automatically fulfills Fermi-liquid sumrules. A rather
complete discussion
up to second order in $U$ has been reported by Menge and M\"uller-Hartmann
\cite{muha89-2,mmuha}. Since it has been pointed out by Georges and Kotliar
\cite{gk}
that these results are not qualitatively much different from the lowest second
order result with Hartree self consistency \cite{yy,zhs}, we shall use the
latter approach here.

The most successful approximate methods for dealing with highly correlated
electron systems
have been developed for the single-impurity Anderson model by choosing the
mixing term as perturbation \cite{kk72}. Unfortunately, the price
one has to pay for leaving the Coulomb interaction intact is that the standard
methods of perturbation theory fail. This problem can be nicely
circumvented for the impurity problem, leading to well defined and controlled
approximation schemes like the so-called NCA \cite{cbw,pg89,kk72}. This
approximation is known to work well when the physics of the system
is dominated by spin fluctuations \cite{gr83,muha84} but fails when charge
excitations become important. In this respect it may be viewed as an approach
complementary to standard perturbation theory. In addition, the NCA tends to
violate Fermi-liquid properties for temperatures much lower than the smallest
energy scale in the problem \cite{muha84}. However, the NCA is nevertheless
quite
reliable over a large interval of parameters including temperature \cite{cbw}.
Since the solution of the model (1) for $d=\infty$ essentially reduces to the
solution of a single-impurity Anderson model it is natural to adopt the NCA
for this problem.

Another natural attempt is to extend the perturbation theory with respect
to mixing directly onto concentrated systems. In this case, however, the
missing features of standard perturbation theory complicate the problem
considerably \cite{gk81} and a controlled approximation (like in the
impurity case) presently does not exist. With the use of some ad-hoc
assumptions it is nevertheless
possible to set up an approximation for this problem, too. These theories are
originally designed for the periodic Anderson model and known in literature
as XNCA \cite{ku85} and LNCA \cite{gr87,gpk88}. Recently, one of the authors
has shown that any such theory for the periodic Anderson model can be readily
employed for the Hubbard model (1), too \cite{gp87,pr91}. In order to obtain
an idea about the quality of these approximations we include the LNCA in
our comparison.

The single-particle density of states  for the Hubbard model (1) at half
filling $n=1$ for several values of $U$ at an inverse temperature $\beta=7.2$
is shown in Fig.~2 for the different kind of approaches discussed before. Let
us first outline the general features of the DOS as they appear from the QMC
results: In all cases one finds two prominent peaks at roughly $\pm U/2$ which
have to be identified with charge excitations on and off the local levels.
In addition there is a pronounced resonance at $\mu$ for small values of $U$
due to coherent movement
of the particles in the system. This feature is suppressed when $U$ is
increased
and eventually a pseudo-gap opens at $\mu$.

In comparing the different approximations to the QMC results the
first thing to note is that the overall agreement between QMC and NCA, apart
from small differences at $\mu$, is very good.
For $U=6$ we did not succeed in  analytically continuing the QMC results. The
only quantity we were able to obtain here is the position of the edges of the
pseudo-gap. These were found to be in good agreement with those predicted by
the NCA. We want to point out that for this value of $U$, as generally
for values $U$ well inside the ``insulator'' phase in Fig.~1 and $\beta U\gg1$,
the NCA does not provide stable results but tends to produce spurious
oscillations at the
gap edges. However, general structures like the width of the pseudo-gap are
reproduced with good accuracy. Nevertheless, these instabilities prohibit a
thorough investigation of this surely very interesting part of the phase
diagram at half filling. We want to emphasize that this problem {\em is
not intrinsic to the NCA}, but rather must be attributed to numerical
instabilities of the computer code used to solve the NCA equations. The reason
is that structures in the NCA equations become very sharp in this region
and eventually cannot be resolved on a discrete energy mesh. When this
occurs we  approximate these structures as poles, which
gives rise to the mentioned numerical instabilities. Note, that this problem
does not occur outside the ``insulator'' phase and off half filling.
A rather interesting point is that ``poles'' in the NCA
begin to develop exactly when the DOS at $\mu$ becomes exponentially
small. This empirical observation was also confirmed by QMC for some
characteristic points in the phase diagram and eventually used to
find an estimate of the right hand border of the crossover region in Fig.~1.

Apparently, perturbation theory in $U$ generally fails to reproduce even
qualitatively
both the high and the low-energy features of the DOS. Only for
small $U$ may the quasiparticle band at $\mu$ be regarded as a good
approximation
to the QMC data. The LNCA on the other hand looks like a too large
value of $U$ had been used. This may be attributed to the approximations
involved
which put a strong emphasis on local correlations and are thus likely to
overestimate residual local interactions.
It also clearly overestimates the charge excitation bands
and shows little of the finer structure near the gap edges \cite{I}, but at
least it reproduces the general features of the DOS qualitatively correctly and
accounts for the existence of the pseudo-gap.

In Fig.~3 we present some typical results off half filling, namely for
$\mu=1$ ($n_e\approx0.94$) and $\mu=0.5$ ($n_e\approx0.8$) at two different
temperatures $\beta=3.6$ and $\beta=14.4$. The value of the Coulomb repulsion
is $U=4$. Again, QMC and NCA are in good agreement
concerning the high and the low energy features except for Fig.~3(d), where
$A_{\rm NCA}(\mu)$ comes out much too large, i.e.\ the NCA fails to account
properly for the low energy physics. This is the principal failure directly
related to the approximations involved in the NCA \cite{muha84}.

Interestingly, the LNCA gives a much weaker temperature
dependence of the DOS at $\mu$, indicating that this approximation
underestimates
the characteristic low energy scale $T_0$.
This is in accordance with the observation made earlier, namely that the LNCA
tends to
overestimate the role of the local correlations. Apart from this failure
the general form of the spectra agrees at least qualitatively with the exact
result.  To obtain reasonable results from perturbation theory, we found
it necessary to fix the occupancy to the QMC value by adjusting
the chemical potential.  This given, the
perturbation theory apparently becomes better with
increasing hole concentration. It nevertheless produces features
which are too broad and rather poor imitations of the charge excitation peaks.

A first conclusion one may draw from these considerations is that
the NCA reproduces most of the general features of the single particle
DOS with good accuracy.
However, Fig.~3(d) clearly shows that for some choice of parameter values
the most important region at $\mu$ is approximated very poorly.
In order to achieve a better classification of the portion of the parameter
space where the NCA constitutes a reliable approximation to the problem
let us substantiate the differences between QMC and NCA by looking at the
quasiparticle weight defined by
\begin{equation}
\zeta^{-1}(T)=1-\frac{{\rm Im}\Sigma(i\omega_0)}{\omega_0}\;\;,
\end{equation}
where $\omega_0=\pi T$ is the lowest Matsubara frequency \cite{serene}.
Figure~4 displays this function for $\mu=1$ and $U=4$ as obtained from QMC
(circles) and NCA (squares).  Note that for these parameter values we expect
$T_0\approx1/8$ \cite{I}, i.e.\ $\zeta^{-1}(T\to0)\approx8$.
The agreement between QMC and NCA is satisfactorily, especially for
temperatures
$T\agt T_0/2$. For temperatures $T\alt T_0/2$ the values obtained by the NCA
become too
large although the order of magnitude is still good. Things become worse as
soon as $T\ll T_0$, where the NCA produces again an upturn instead of a
saturation. Both,
the slightly too large values as well as the failure for $T\alt T_0/5$, must
be attributed to the well known pathology of the NCA \cite{muha84}. In our
case,
the most important aspect of this pathology is its tendency to give a slightly
too small absolute value for the self energy near $\mu$ \cite{cbw}.
While for the Anderson
impurity model this behaviour is not important for temperatures $T\agt0.1T_K$
\cite{cbw},
the self-consistency process involved here naturally accumulates
this deficiency, leading to the observed small discrepancies between QMC and
NCA
in the spectra and $\zeta^{-1}(T)$. Eventually the NCA breaks down completely
for temperatures small compared to $T_0$. However, from our comparison one
can
conclude that the results produced by the NCA are reliable for $T/T_0\agt1/5$.

For small $U$ and/or far away from half filling the latter restriction makes
the NCA obviously rather useless, because the low temperature energy scale is
usually of the order $O(1)$ here (see, for example, Fig.~3d with $\mu=1.5$
and $\beta=14.4$).
However, in the interesting region of large $U$ and close to half filling the
low temperature scale $T_0$ is much smaller. In these cases the NCA provides a
fast and consistent way to obtain information that is hard to access by other
methods.
For example, let us discuss the one-particle self energies and the reentrance
behaviour found in the phase diagram at half filling (see Fig.~1).

The imaginary part of the one-particle self energy is a quantity interesting
in its own right
since it provides valuable information about the low temperature
behaviour of the system. For a normal Fermi liquid when $T\to0$, one expects
$-{\rm Im}\Sigma(\omega+i\delta)$ to exhibit a parabolic minimum at $\mu$ with
a curvature and
temperature dependence that is
characteristic of the effective mass of the quasiparticles in the system.
A way to obtain this latter information has already been discussed with the
definition
of the quasiparticle weight equation (8). Figure 5 gives an impression of how
the self energy behaves for some parameter values, namely at half filling
($n_e=1$) for two values of $U=2,\ 4$
for a fixed temperature $\beta=7.2$ in
Fig.~5(a) and off half filling ($n_e\approx0.94$), for a fixed $U=4$ and
two characteristic temperatures $\beta=3.6,\;28.8$ in Fig.~5(b).
While for $U=2$ (full curve in Fig.~5a) and off half filling (Fig.~5b)
$|{\rm Im}\Sigma|$ obviously develops a nice parabolic minimum at $\mu$, the
behaviour for $U=4$ at half filling (broken curve in Fig.~5a) is completely
different. Here a sharp peak at $\mu$ appears separated by a (pseudo-)
gap from the continuum of particle-hole excitations.
{}From general arguments \cite{muha89-2}
it follows that $-{\rm Im}\Sigma(\mu+i\delta)\approx1/A(\mu)$ in this case.
It is clear that
one will never obtain a Fermi liquid with this type of self energy
\cite{comment}.
Physically, this peak corresponds to an effective resonant scattering provided
by the medium surrounding a given particle, thus localizing it by forming a
bound
state. It is nevertheless surprising to find such a structure when general
phase
space arguments rather suggest that particle-hole scattering near $\mu$ has to
vanish \cite{lutt}.
Thus an important question
is whether this structure is stable or may be replaced by the usual minimum
for $T\to0$. This leads us directly to the reentrance behaviour seen in Fig.~1.

To study this interesting behaviour more closely
we fix the Coulomb parameter at $U=3.5$ and scan the temperature from above
the MI-crossover region ($T=0.32$) down to $T=0.002$. Obviously, such a low
temperature cannot be reached with QMC for this value of $U$. The results are
shown in Fig.~6.
One nicely sees the opening of the pseudo-gap as the temperature is lowered.
Eventually, this pseudo-gap is
destroyed by a very narrow resonance at $\mu$ which also signals the onset
of Fermi-liquid behaviour. From the value of $A(\mu)$ we extrapolate to
a low temperature scale $T_0\alt1/400$.  Thus, the Fermi liquid that eventually
emerges has extremely large Fermi-liquid parameters. Another question is why
such an Abrikosov-Suhl resonance can built up from an insulator at all? Here
we must keep in mind that we merely observe a pseudo-gap, i.e.\ the DOS
around $\mu$ is never exactly zero and consequently will lead to a small
but finite low temperature scale $T_0$. Whether the Fermi-liquid phase will
win depends entirely on the balance between the energy gain due to the
delocalization of the particles in the narrow band at $\mu$ and the loss
in correlation energy for the same reasons.  It definitely seems more
favourable for $3<U<4$ but we cannot decide from the data available whether
the transition line will finally intercept the abscissa at a
$U<\infty$ or not. We must stress at this point that this whole scenario is
valid {\em if and only if}\/ we have sufficient magnetic frustration to
suppress or destroy the antiferromagnetic transition appearing in the phase
diagram.

\section{Thermodynamic and Transport properties}
The peculiar features of the single-particle DOS and self energy
discussed in the previous
section motivate a closer inspection of thermodynamic and transport
properties of the Hubbard model (1) in the paramagnetic phase. Except for the
one-dimensional model,
a thorough study of these quantities in the thermodynamic limit was not
possible
yet. Previous results from QMC simulations \cite{hirsch} are usually restricted
to relatively small values of $U$ or comparatively high temperatures. Since
they are carried out on a finite lattice with a discrete energy
spectrum, they probably will also miss the Kondo effect if it persists
in three dimensions.
The simplifications arising in $d=\infty$, however, make it possible to give
closed
expressions for several quantities including the free energy, internal energy
and optical conductivity which involve only the one-particle propagators
in a simple way. We are thus in principle in a position to calculate these
quantities exactly or, since we shall use the NCA to solve equations (2)-(3),
obtain at least a very good approximation for them.

Although the derivation of the expressions for those quantities is straight
forward
we will just state the final results and leave the mathematics to the
appendix.
To start with, the thermodynamic potential $\Omega(T)$ is given by \cite{janis}
\begin{equation}
\beta\Omega
\displaystyle=N\beta\Omega_{imp}-\sum_{{\bf k}}{\rm
Tr}\ln\left(\frac{G_{ii}}{G_{{\bf k}}}\right)\;\;.
\end{equation}
Here, $\Omega_{imp}$ is the local free energy contribution from the effective
Anderson impurity problem.
Although, in principle,  the knowledge of $\Omega(T)$ provides everything one
needs, it is helpful to have an independent expression for the internal energy
$E(T)$, too. The main reason is that thermodynamic quantities are usually
obtained by differentiating $\Omega(T)$, which is a rather unpleasant task from
a numerical point of view. In particular, the specific heat is a second
derivative of $\Omega$, but it is a first of $E(T)$.
An expression for $E(T)$ is given by
\begin{equation}
E(T)=\frac{1}{2}\sum_{{\bf k}\sigma}\int d\omega f(\omega)\left(
t_{\bf k}+\omega\right)A_{{\bf k}\sigma}(\omega)+\frac{1}{2}\mu N_e\;\;\;.
\end{equation}

The last quantity we want to study in this paper is the conductivity.
We restrict ourselves to the ${\bf q}={\bf 0}$ component, because without
coupling
to elastic degrees of freedom we do not expect the model (1) to exhibit any
incommensurate charge density instability, i.e.\ the ${\bf q}={\bf 0}$
component will be the most important one.
In this particular case,
the limit $d\to\infty$ provides us with an extreme simplification, namely
one can easily show (see e.g.\ ref.\ 28 and appendix) that the expression for
$\sigma(\omega,T)$ reduces to
\widetext
\begin{equation}
\sigma(\omega)=\pi \int d\omega'\int d\epsilon
A_0(\epsilon)A(\epsilon,\omega')A(\epsilon,\omega'+\omega)
\frac{f(\omega')-f(\omega'+\omega)}{\omega}\;\;.
\end{equation}
\narrowtext
Note that for $\omega\to0$ this is very similar to the result of Schweitzer
and Czycholl \cite{sc91}, except that our result is written as energy integrals
and thus avoids their explicit sum on lattice sites that is impossible
to evaluate in $d\to\infty$.

Let us begin with a discussion of the properties of the model (1) at
half filling. Due to numerical difficulties the NCA is currently not able to
provide stable enough results in the interesting region just above the
``MI''-crossover line. We therefore have to concentrate on a value just
below the critical one and we found it to be a convenient choice to use $U=3$.
As it turns out, the behaviour found here is already close
to what one may expect in the ``insulating'' region.

Before we turn to the actual thermodynamic properties, we first want to give
with Fig.~7 an impression of the variation of the one-particle spectra
with temperature.
It is clear that the dip in the DOS at $\mu$ for higher temperatures
is a poor replacement for the actual
exponentially small DOS at larger values of $U$. However, together with the
Abrikosov-Suhl resonance at low temperatures it gives a fairly good picture
of the general temperature dependence of the DOS even for $U>U_c$. From it we
may anticipate the behaviour of the various thermodynamic and transport
properties: Starting from high temperatures, one will encounter a
temperature regime (e.g.\ $t^*>T\gg T_0$) where the DOS mainly consists of two
separated bands.
For the entropy for example this means that it will be rather flat with a value
reflecting the degeneracy of the states in the lower band, i.e.\ $S\approx\ln
2$.
At the same time the specific heat will decrease and become very small.
If there
were a true gap we would actually expect $C_V\sim\exp(-\beta\Delta_{Gap})$. The
resistivity,
on the other hand will be large and increases with decreasing temperature,
while
the optical conductivity is governed by the charge excitations of energy $U$
and shows no Drude peak.

Figure 8 displays the different thermodynamic quantities for the parameters
under consideration as a function of temperature. In addition, we include for
comparison some
values of the internal energy $E(T)$ as obtained from QMC (circles). Again, QMC
and NCA are
found to be in good agreement. It is noteworthy, that the internal energy
becomes rather flat at $T\approx1/5$. At the same time the
entropy has a saddle point with a value of $S\approx\ln2$, and the
specific heat becomes small, as expected. The unphysical variation of $S(T)$
and $C(T)$ found in this interval must be attributed to
numerical inaccuracy.
In fact, by increasing the precision of the results we observed that e.g.\ the
nonmonotonic variation in $S(T)$ is reduced considerably while the value seems
to approach $S\approx\ln2$ with good accuracy. Also, the internal energy
appears
to be much more insensitive to numerical inaccuracies than the free energy.

When the temperature is further lowered,
we see a decrease of $S(T)$ again, accompanied by a strong increase of $C(T)$,
which eventually shows a maximum. This peak in $C(T)$ is a further fingerprint
of the
Kondo-effect in this model. Unfortunately, a further decrease in temperature
is not possible, since the NCA-pathology becomes important for $T\alt0.01$.
{}From our experience this points towards a proper low temperature energy
scale of about $T_0\approx 1/20$, which is also consistent with the position
of the peak in $C(T)$ at $T\approx T_0/3$.
At present one can only infer from the knowledge of the properties of heavy
fermion materials that $C(T)\sim T/T_0$ for temperatures below the maximum.

{}From these results one may easily extrapolate to the behaviour of the system
for $U>3$.
In fact, we mainly expect the extent of the flat
region in $E(T)$ and $S(T)$ to become larger, namely it should last roughly
until a possible crossover into the Kondo regime begins. Then one will find
a very steep decrease of both quantities again. For values of $U$ smaller
than $U=3$ on the other hand the flat region will shrink and the slope of the
decrease for lower
temperatures will become smaller.

The resistivity for these parameter values as a function of temperature is
shown in Fig.~9. Consistent with the DOS and the thermodynamics we first
observe
a semi-metallic increase at high temperatures which goes through a maximum and
then
decreases for low temperatures. Since we expect the system to behave like a
Fermi liquid at low temperatures, the resistivity should follow
$\rho(T)= \rho_0+a\cdot(T/T_0)^2$, where $\rho_0=0$, $T_0\approx1/20$ and
$a=O(1)$. The calculated
data generally follow this law when we also allow for a small intercept
$\rho_0\approx-5\cdot10^{-2}$ (see inset in Fig.~9). It is obvious that this
negative intercept is a pure artifact and related to the pathology of the NCA:
As discussed earlier, the
approximations in the NCA tend to give a too small value of
${\rm Im}({\cal G}^{NCA}(\omega+i\delta))^{-1}$
near $\mu$. Since the self-energy is given by the difference between this
quantity and the effective hybridization eq.\ (3), one will eventually
encounter a temperature where causality is violated and the results by the NCA
become meaningless. For the present parameter values this happens for
$T\alt1/100$. It is clear, that this breakdown will manifest itself strongest
right at the minimum of ${\rm Im}\Sigma(\omega-i\delta)$, which happens to
be always exactly at $\mu$ at half filling due to the particle-hole symmetry.
Since the low-temperature resistivity on the other hand is approximately just
given by ${\rm Im}\Sigma(-i\delta)$ \cite{mura}, this violation of causality
produced by the NCA leads directly to the observed  unphysical value of
$\rho_0$.

The picture for half filling is completed by the optical conductivity in
Fig.~10. As expected,
the case $U=3$ already gives an idea how the ``insulator'' will look like:
For high temperatures one finds a weak vestige of the Drude peak for
$\omega\to0$ which at first is suppressed when the temperature is
lowered. Note, however, that we always maintain a finite value for $\sigma(0)$
consistent with the DOS in Fig.~7. At the same time the spectral weight of the
charge excitation peak at $\omega=U$ increases. When we further lower the
temperature,
the situation reverses. A Drude peak at $\omega=0$ builds up again (see inset
in Fig.~10)
and the spectral
weight at $\omega=U$ is decreased. In addition, a shoulder emerges at
$\omega\approx1$.
This feature must be ascribed to the additional excitations from the lower
Hubbard band to the Abrikosov-Suhl resonance at $\mu$. Note that the spectral
weight
associated with this additional resonance is rather small.

Let us finish the discussion of the half filled case with a look at the typical
behaviour of the optical conductivity inside the ``insulator'' region in
Fig.~1.
Although the numerical problems prohibit a discussion of thermodynamic
quantities
here the results prove to be stable enough to allow the calculation
of $\sigma$. Figure 11 shows our
results for the DOS (Fig.~11(a)) and $\sigma(\omega)$ (Fig.~11(b)) for a value
of $U=6$ and temperatures $\beta=1.44$ (full curves) and $\beta=28.8$ (broken
curves). The high temperature result represents a point in the semi-metallic
portion of the phase diagram (Fig.~1) and still shows a small peak at
$\omega=0$ due
to the thermally induced states in the gap here. This feature is
however completely lost in
the ``insulating phase'', where only the charge excitation peak at $\omega=U$
survives. We point out that the high temperature results for $U=3$
are indeed similar to the general behaviour found here even though the larger
DOS
at $\mu$ of course leads to a finite value for $\sigma(0)$ there.

The situation off half filling is studied for the case $U=4$ and $n_e=0.97$.
The variation of the DOS with temperature for these parameter
values is collected in Fig.~12. Compared to the temperature dependence
of $A(0)$ in Fig.~7, we observe a slower increase here, i.e.\ we have a
somewhat
smaller low-temperature scale $T_0$. From the value of $A(0;T=1/28.8)\approx
2/\sqrt{\pi}$ we extrapolate
to a $T_0\alt 1/30$ for these parameter values. Although we are in principle
able to trace the
properties of the model (1) for a fixed electron density, it turns out that the
obtainable numerical accuracy is not sufficient to get a reasonable result for
the free energy. However, as already mentioned for the half filled case, the
internal energy $E(T)$ is much more well behaved and we shall
concentrate on its behaviour here.
The results for the thermodynamics are collected in Fig.~13. The features
found are actually very similar to the ones known from heavy fermion physics,
as expected: We observe a maximum in $C(T)$ at approximately $T_0/3$, where
$T_0\approx 1/30$ was read off the half-height of the DOS in Fig.~12. Note
that in contrast to half filling we do not have a pronounced flat region in
$E(T)$ or $S(T)$ here.
The values for the entropy in Fig.~13 were obtained by direct integration
$S(T)=\int C(T)/TdT$. Taking into account the decreasing relative precision
in $E(T)$ and the resulting large errors in $C(T)$ for lower temperatures
we find, as expected, an entropy $S\approx\ln2$ associated with the low
temperature peak
in $C(T)$.

The resistivity in this case is shown in Fig.~14. As in the picture for half
filling we
observe an increase of $\rho(T)$ at high temperatures which
eventually goes through a maximum. For low temperatures we find
a power
law $\rho(T)=a\cdot(T/T_0)^2$ with $T_0\approx 30$ and $a=O(1)$ consistent with
Fermi liquid theory.
This time we do not observe any unphysical behaviour
down to the lowest temperatures studied. This may be related to the fact,
that the slight shift of the minimum of ${\rm Im}\Sigma(\omega-i\delta)$
above $\mu$ sufficiently reduces the influence of the NCA-pathology here.

It is noteworthy that for half filling and off half filling the position of
the maximum in $\rho(T)$ does not seem to be related to $T_0$ in a way
similar to heavy-fermion systems. In fact, from the position
of these maxima one would rather tend to rate the systems as weakly correlated.
It is, however, important to remember that since the DOS at $\mu$ is strongly
temperature dependent, the Kondo scale itself is a function of temperature.

Finally we present the optical conductivity for the parameter values $U=4$ and
$n_e=0.97$ in Fig.~15 for some characteristic temperatures. The general
structure
is similar to half filling except that the Drude peak for $\omega
\to 0$ continuously develops when the temperature is decreased. Also, the
``Kondo''-shoulder found in Fig.~10 is not
visible here. Again, the weaker temperature dependence of $\sigma(\omega\to0)$
points towards a smaller $T_0$ in this case.

\section{Summary}
In this paper we have presented a detailed study of thermodynamic and
transport properties of the Hubbard model on a infinite-dimensional hypercubic
lattice. In contrast to the previous study \cite{ja92,I} we did not take into
account the antiferromagnetic ordering expected for this model with nearest
neighbour transfer,
but concentrated on the paramagnetic case. The importance of such a study is
motivated by the fact that inclusion of transfer beyond nearest neighbours will
magnetically frustrate the system and thus depress the ordering. Besides, the
effect of an antiferromagnetic transition on thermodynamic quantities like
$S(T)$ and $C(T)$ is well known once their behaviour in the paramagnetic regime
is known.

The first part of this paper was devoted to a comparison between different
approaches to the Hubbard model. As reference point we used the essentially
exact QMC method discussed extensively in refs. 5 and 7. We found that a
good description can be achieved by using the NCA to approximately
solve the impurity Anderson model which enters the solution of (1) in infinite
dimensions. Using the NCA, the self-consistent set of equations can be
solved very quickly which enabled us to present a variety of quantities
as functions of temperatures for physically meaningful parameters.
One of our main results is the interesting variation of the entropy and
specific
heat in the half filled case. We believe that this behaviour can be viewed
as generic for the strongly correlated model. This conjecture was basically
confirmed by the qualitative similarity between our results for half filling
and
$n=0.97$.

Unfortunately, the NCA-approach breaks down for too low temperatures
due to an intrinsic violation of Fermi-liquid properties. However, our
low-temperature results strongly suggest that a heavy-electron liquid builds up
with a unique energy scale deductable from the variation of the various
physical quantities. One thus can in principle adopt the well
developed phenomenology for these systems to extrapolate to a consistent
low-temperature limit for the paramagnetic phase  of the infinite dimensional
Hubbard model \cite{ohkawa}.

Together with our previous study, to the extent of our knowledge this
represents the first consistent and reliable study of dynamic and
thermodynamic properties of the $d>1$ Hubbard model
in the thermodynamic limit. Although the qualitative effects of finite
dimensional corrections are not well understood, we believe that many of
the features found here will basically persist in at least $d=3$.
\acknowledgements

This research was supported by the National Science Foundation grant number
DMR-88357341, the National Science Foundation grant number DMR-9107563, the
Ohio State University center of materials research and by the Ohio
Supercomputing
Center.

\appendix{Expressions for $\Omega(T)$ and $E(T)$}
In the following we will provide the derivation of the expressions for
the free energy (9) and internal energy (10) for the Hubbard model in
$d\to\infty$.

Let us start with the free energy.
According to Baym \cite{janis,ba62} one can quite generally write the grand
canonical
potential as
\begin{equation}
\beta\Omega(\underline{\bf\Sigma})=\Phi(\underline{\bf G})-{\rm Tr}(
\underline{\bf\Sigma}\cdot\underline{\bf G})
-{\rm Tr}\ln((\underline{\bf G}^0)^{-1}-\underline{\bf\Sigma})
\end{equation}
where shorthand matrix notations
$$(\underline{\bf G})_{ij;nm}=G_{ij}(i\omega_n,i\omega_m)$$
and
$$(\underline{\bf\Sigma})_{ij;nm}=\Sigma_{ij}(i\omega_n,i\omega_m)$$
were introduced.
The functional $\Phi$ is defined via the perturbation expansion
of $\underline{\bf\Sigma}$ in terms of $\underline{\bf G}$ by the property
\begin{equation}
\frac{\delta\Phi}{\delta\underline{\bf G}}=\underline{\bf\Sigma}\;\;.
\end{equation}
For $d\to\infty$
we know that $\underline{\bf\Sigma}(z)\equiv\Sigma(z)\underline{\bf 1}$ and its
perturbation expansion
involves only the local component of $\underline{\bf G}$, $G_{ii}(z)$. This
implies that
$\Phi(\underline{\bf G})=\Phi(\{G_{ii}\})=N\Phi_{imp}(G_{ii})$ \cite{bm,janis}
and finally
\widetext
\begin{equation}
\begin{array}{l@{\;=\;}l}\displaystyle
\beta\Omega(T) & \displaystyle N\Phi_{imp}(G_{ii})
-N{\rm Tr}(\Sigma G_{ii})-N{\rm Tr}\ln(G_{ii}^{-1})
-\sum_{{\bf k}}{\rm Tr}\ln\left(\frac{G_{ii}}{G_{{\bf k}}}\right)\\[5mm]
&\displaystyle N\beta\Omega_{imp}-\sum_{{\bf k}}{\rm
Tr}\ln\left(\frac{G_{ii}}{G_{{\bf k}}}\right)\;\;.
\end{array}
\end{equation}
\narrowtext
Here, $\Omega_{imp}$ is the local free energy contribution from the effective
Anderson impurity problem for a given $\Sigma$ \cite{janis}.

For the derivation of $E(T)$ we first note, that from
$\Omega=E-TS-\mu N$ one obtains for fixed particle number
$E=\partial(\beta \Omega)/\partial\beta+\mu N+\beta\partial\mu/\partial\beta
N_e$. With $\Omega=-\beta^{-1}\ln Z$
this leads to the relation $E(T)=\langle H\rangle+\mu N_e$. Note
that this unusual form arises because our definition of the Hubbard Hamiltonian
(1) absorbs the term $-\mu N$.
In order to calculate the expectation value, let us evaluate
the following commutator \cite{grewe}:
\widetext
\FL\begin{equation}
\begin{array}{l@{\;=\;}l}
\displaystyle\sum_{i\sigma} c^\dagger_{i\sigma}[c_{i\sigma},H] &
\begin{array}[t]{l}
\displaystyle\sum_{i,l,j,\sigma}t_{jl}(c^\dagger_{i\sigma}[c_{i\sigma},
c^\dagger_{l\sigma}c_{j\sigma}]+c^\dagger_{i\sigma}[c_{i\sigma},
c^\dagger_{j\sigma}c_{l\sigma}])
-\mu\sum_{ij\sigma}c^\dagger_{i\sigma}[c_{i\sigma},n_{j\sigma}]\\[5mm]
\displaystyle+\sum_{i,l,\sigma}c^\dagger_{i\sigma}[c_{i\sigma},
n_{l\uparrow}n_{l\downarrow}]\end{array}\\[10mm]
 &  H_{kin}-\mu N+2U = 2H-H_{kin}+\mu N\;\;.
\end{array}
\end{equation}
That means for $\langle H\rangle$
\FL\begin{equation}
\begin{array}{l@{\;=\;}l}
\displaystyle\langle H\rangle&
\displaystyle\frac{1}{2}\langle H_{kin} +
\sum_{i\sigma}c^\dagger_{i\sigma}[c_{i\sigma},H]\rangle-\frac{1}{2}
\mu N_e \\[5mm]
&\displaystyle
\frac{1}{2}\sum_{{\bf k}\sigma} t_{\bf k}\langle c^\dagger_{{\bf k}\sigma}
c_{{\bf k}\sigma}\rangle
+\frac{1}{2}\sum_{i\sigma}\langle c^\dagger_{i\sigma}[c_{i\sigma},H]
\rangle-\frac{1}{2}\mu N_e\\[5mm]
 &\displaystyle\frac{1}{2}\sum_{{\bf k}\sigma} t_{\bf k}G_{{\bf
k}\sigma}(-\delta)
+\frac{1}{2}\sum_{i\sigma}G_{[c_{i\sigma},H],c^\dagger_{i\sigma}}(-\delta)
-\frac{1}{2}\mu N_e\\[5mm]
 &\displaystyle\frac{1}{2\beta}\sum_{i\omega_n}\left(\sum_{{\bf k}\sigma}
t_{\bf k}G_{{\bf k}\sigma}(i\omega_n)
+\sum_{i\sigma}G_{[c_{i\sigma},H],c^\dagger_{i\sigma}}(i\omega_n)\right)
e^{i\omega_n\delta}-\frac{1}{2}\mu N_e
\end{array}
\end{equation}
With the equation of motion
\begin{equation}
zG_{ii}(z)=1+G_{[c_{i\sigma},H],c^\dagger_{i\sigma}}(z)
\end{equation}
and using translational invariance, i.e.\ $G_{ii}(z)=\frac{1}{N}\sum
G_{{\bf k}\sigma}(z)$,
one arrives at
\FL\begin{equation}
E(T)=\frac{1}{2\beta}\sum_{i\omega_n}\sum_{{\bf k}\sigma}\left(
t_{\bf k}G_{{\bf k}\sigma}(i\omega_n)+i\omega_nG_{{\bf k}\sigma}(i\omega_n)-1
\right)e^{i\omega_n\delta}+\frac{1}{2}\mu N_e\;\;\;.
\end{equation}
Now,
$$
\frac{1}{\beta}\sum_{i\omega_n}G_{{\bf k}\sigma}(i\omega_n)=\int d\omega
f(\omega)
A_{{\bf k}\sigma}(\omega)
$$
and with $\int d\omega A_{{\bf k}\sigma}(\omega)=1$
\widetext
$$
\begin{array}{l@{\;=\;}l}
\displaystyle\frac{1}{\beta}\sum_{i\omega_n}
\left(i\omega_nG_{{\bf k}\sigma}(i\omega_n)-1\right)e^{i\omega_n\delta} &
\displaystyle\frac{1}{\beta}\sum_{i\omega_n}\int d\omega A_{{\bf
k}\sigma}(\omega)\left(
\frac{i\omega_n}{i\omega_n-\omega}-1\right)e^{i\omega_n\delta}\\[5mm]
& \displaystyle\frac{1}{\beta}\sum_{i\omega_n}\int d\omega A_{{\bf
k}\sigma}(\omega)
\frac{\omega}{i\omega_n-\omega}e^{i\omega_n\delta}\\[5mm]
&\displaystyle\int d\omega\omega f(\omega) A_{{\bf k}\sigma}(\omega)
\end{array}
$$
\noindent the final equation reads
\FL\begin{equation}
E(T)=\frac{1}{2}\sum_{{\bf k}\sigma}\int d\omega f(\omega)\left(
t_{\bf k}+\omega\right) A_{{\bf k}\sigma}(\omega)+\frac{1}{2}\mu N_e\;\;\;.
\end{equation}
\narrowtext
\appendix{Optical conductivity}
In general, the conductivity tensor is expressed via the current-current
susceptibility as
\begin{equation}
\sigma_{kl}(\omega)\begin{array}[t]{l}
\displaystyle=-\frac{1}{N}{\rm Re}\left\{
\frac{1}{i\omega}\chi_{j_l,
j_k}(\omega+i\delta)\right\}\\[5mm]
\displaystyle=\frac{1}{N}{\rm Re}\left\{\frac{1}{i\omega}\langle\langle
j_l|j_k\rangle\rangle(\omega+i\delta)\right\}\\[5mm]
\displaystyle\equiv{\rm Re}\tilde\sigma_{kl}(\omega+i\delta)\;\;.
\end{array}
\end{equation}
For the simple cubic lattice under consideration, the tensor is proportional to
the unit tensor, i.e.
\begin{equation}
d\tilde\sigma(z)=\frac{1}{N}\frac{1}{iz}\sum_l\langle\langle
j_l|j_l\rangle\rangle(z)\;.
\end{equation}
Further, the current operator for the Hubbard model (1) is given by
($e=\hbar=1$)
\begin{equation}
{\bf j}=\sum_{{\bf k}\sigma}{\bf v}_{\bf k}n_{{\bf k}\sigma}
\end{equation}
where the group velocity or the particles is defined via
${\bf v}_{\bf k}={\bf\nabla} t_{\bf k}$. This leads to
\FL\begin{equation}
d\tilde\sigma(z)=\frac{1}{iz}\frac{1}{N}\sum_{{\bf k},{\bf k}',\sigma}
\sum_lv_{k_l}v_{k'_l}\langle\langle n_{{\bf k}\sigma}|n_{{\bf k}'\sigma}
\rangle\rangle(z)\;\;.
\end{equation}
This expression has the perturbation expansion shown in Fig.~16. One important
implication of the limit $d\to\infty$ is, that the irreducible vertex
$\Gamma(i\omega_n,i\omega_m;i\nu)$ in Fig.~16
has to be purely local \cite{muha89-1}. This means that the ${\bf
k}$-summations
in the second part of Fig.~16 can be performed independently and thus these
vertex
corrections vanish, because ${\bf v}_{\bf k}$ has a different parity than
$t_{\bf k}$
\cite{khurana}. This means, that only the simple bubble survives and we are
left with (the lattice constant to be taken unity)
\widetext
\FL\begin{equation}
\tilde\sigma(i\nu)
=\frac{1}{\nu}\frac{1}{N\beta}\sum_{{\bf k}\sigma,\omega_n}\sum_l
v^2_{k_l}G_{{\bf k}\sigma}(i\omega_n)G_{{\bf k}\sigma}(i\omega_n+i\nu)
\displaystyle=\frac{1}{\nu}\frac{1}{N\beta}\sum_{{\bf k}\sigma,\omega_n}\sum_l
4t^2\sin^2(k_l)G_{{\bf k}\sigma}(i\omega_n)G_{{\bf k}\sigma}(i\omega_n+i\nu)\
\end{equation}
for nearest-neighbour transfer.

\narrowtext
The problem left is to evaluate the sum
\begin{equation}
\tilde{\rho}_0(\epsilon)=\frac{1}{N}\sum_{\bf k}\sum_l
\sin^2(k_l)\delta(\epsilon-t_{\bf k})
\end{equation}
For this purpose, we use the method applied by M\"uller-Hartmann
\cite{muha89-1} and study the Fourier-transform of (B6):
\widetext
\FL\begin{equation}
\Psi_d(s)\begin{array}[t]{l}
\displaystyle=\int\tilde{\rho}_0(\epsilon)e^{is\epsilon}d\epsilon
=\frac{1}{N}\sum_{\bf k}\sum_l\sin^2(k_l)e^{ist_{\bf k}}\\[5mm]
\displaystyle=
d\left[\int_{-\pi}^\pi e^{-2ist\cos k}\frac{dk}{2\pi}\right]^{d-1}
\left[\int_{-\pi}^\pi\sin^2k e^{-2ist\cos k}\frac{dk}{2\pi}\right]
=\frac{d}{2}\left[J_0(2st)\right]^d+\frac{d}{2}\left[J_0(2st)\right]^{d-1}
J_2(2st)\;\;,
\end{array}
\end{equation}
where $J_\nu(z)$ are Bessel functions. Noting that $A_0(\epsilon)
=\int dse^{-is\epsilon}\left[J_0(2st)\right]^d$, we find
\begin{equation}
\frac{1}{N}\sum_{\bf k}\sum_l\sin^2(k_l)\delta(\epsilon-t_{\bf k})=
\frac{d}{2}\left(\rho_0(\epsilon)+\int_{-\infty}^\infty ds e^{-is\epsilon}
\left[J_0(2st)\right]^{d-1}J_2(2st)\right)\;\;.
\end{equation}
\narrowtext
This relation is valid for a simple cubic lattice and nearest neighbour
transfer
in any dimension. For the current purpose we are
only interested in the limit $d\to\infty$, where one can achieve a further
substantial simplification. Noting that $2t\sim1/\sqrt{d}\ll1$ we may
approximate $J_2(2st)\approx s^2t^2$ for large $d$ and find for the last term
in (B8)
\FL\begin{equation}
\int dse^{-is\epsilon}\left[J_0(2st)\right]^{d-1}J_2(2st)
\begin{array}[t]{l}\displaystyle\approx
t^2\int ds s^2e^{-is\epsilon}\left[J_0(2st)\right]^d\\[5mm]
\displaystyle=-t^2\frac{d^2A_0(\epsilon)}{d\epsilon^2}\;\;.\end{array}
\end{equation}
Since $t^2\sim1/d$ this term is negligible in the limit $d\to\infty$, i.e.\
$\tilde{\rho}_0(\epsilon)=d\cdot\rho_0(\epsilon)/2$ and finally
\widetext
\FL\begin{equation}
\tilde\sigma(i\nu)
=\frac{1}{\nu}\frac{1}{\beta}\sum_{\omega_n}
\int d\epsilon A_0(\epsilon)G(\epsilon,i\omega_n)G(\epsilon,i\omega_n+
i\nu)
\displaystyle=\frac{1}{\nu}\int d\epsilon d\omega d\omega'A_0(\epsilon)
A(\epsilon,\omega)A(\epsilon,\omega')
\frac{f(\omega)-f(\omega')}{\omega-\omega'+i\nu}\;\;.
\end{equation}
Taking the real part of the analytic continuation of equation (B10) leads to
our final result for the optical conductivity
\widetext
\begin{equation}
\sigma(\omega)=\pi \int d\omega'\int d\epsilon A_0(\epsilon)
A(\epsilon,\omega')A(\epsilon,\omega'+\omega)
\frac{f(\omega')-f(\omega'+\omega)}{\omega}\;\;.
\end{equation}
\narrowtext
When we collect the missing constants we find for the unit of the conductivity
$$
\sigma_0=\frac{\pi e^2a^2{t^*}^2}{2\hbar}\frac{N}{Vol}\approx
10^{-2}\ldots10^{-3} [\mu\Omega cm]^{-1}\;\;,
$$
for $t^*\approx 1{\rm eV}$ and $a=O(a_0)$.


\begin{references}
\bibitem{comment} This particular behaviour also explains why any finite order
perturbation theory in $U$ will have serious problems to recover this result,
since these theories produce Fermi-liquid behaviour by construction
\cite{lutt}.
\bibitem{grewe} This nice trick was pointed out to us by N.\ Grewe.
\end{references}
\end{document}